# Glycerol Modulates Water Permeation through *Escherichia coli* Aquaglyceroporin GlpF

Liao Y. Chen

Department of Physics, University of Texas at San Antonio, One UTSA Circle, San Antonio, Texas 78249 USA

**ABSTRACT**

Among aquaglyceroporins that transport both water and glycerol across the cell membrane, *Escherichia coli* glycerol uptake facilitator (GlpF) is the most thoroughly studied. However, one question remains: Does glycerol modulate water permeation? This study answers this fundamental question by determining the chemical-potential profile of glycerol along the permeation path through GlpF's conducting pore. There is a deep well near the Asn-Pro-Ala (NPA) motifs (dissociation constant 14 µM) and a barrier near the selectivity filter (10.1 kcal/mol above the well bottom). This profile owes its existence to GlpF's perfect steric arrangement: The glycerol-protein van der Waals interactions are attractive near the NPA but repulsive elsewhere in the conducting pore. In light of the single-file nature of waters and glycerols lining up in GlpF's amphipathic pore, it leads to the following conclusion: Glycerol modulates water permeation in the µM range. At mM concentrations, GlpF is glycerol-saturated and a glycerol dwelling in the well occludes the conducting pore. Therefore, water permeation is fully correlated to glycerol dissociation that has an Arrhenius activation barrier of 6.5 kcal/mol. Validation of this theory is based on the existent *in vitro* data, some of which have not been given the proper attention they deserved: The Arrhenius activation barriers were found to be 7 kcal/mol for water permeation and 9.6 kcal/mol for glycerol permeation; The presence of up to 100 mM glycerol did not affect the kinetics of water transport with very low permeability, in apparent contradiction with the existent theories that predicted high permeability (0 M glycerol).

## INTRODUCTION

*Escherichia coli* aquaglyceroporin GlpF is a member of the membrane proteins responsible for water and solute transport across the cell membrane [1-6]. Among the aquaglyceroporin sub-family of proteins that conduct both water and glycerol, GlpF is the most thoroughly studied, both *in vitro* [7-19] and *in silico* [18, 20-29]. There is no controversy over the science that GlpF conducts both water and glycerol and how the amphipathic pore of GlpF selectively facilitates the passage of waters and glycerols lining up in a single file through the conducting channel [17, 18, 20, 21]. However, one fundamental question remains: Does glycerol modulate water permeation through GlpF? And, related to this question, there are some unsolved issues about water permeation through this protein's conducting pore: The *in vitro* data indicate that GlpF is much less permeable to water than *Escherichia coli* aquaporin Z (AQPZ) and other water-selective aquaporins [13, 14, 30], but theoretical studies predict that GlpF is more permeable than AQPZ *etc*. [23, 31]; The *in vitro* experiments show that water permeation has an Arrhenius activation barrier that is about 7 kcal/mol [13], but the theoretical studies all give a rather flat free-energy profile throughout the permeation channel of GlpF[20, 27]. While the *in vitro* experiments indicate that the presence of up to 100 mM glycerol does not affect the kinetics of water transport [13], all *in silico* studies are limited to 0 M glycerol concentration. All these problems can be resolved once we have an accurate determination of the chemical potential of glycerol as a function of its center-of-mass coordinates along a path leading from the periplasm to the entry vestibule of GlpF, through the channel, to the cytoplasm. This chemical-potential profile, considered on the basis of the structure information available in the literature [17, 18], can ascertain the conclusion that glycerol strongly modulates water permeation through GlpF.

Inside the GlpF channel, waters and glycerols line up in a single file, occluding one another from occupying the same z-coordinate. (The z-axis is chosen as normal to the membrane-water interface, pointing from the periplasm to the cytoplasm.) Therefore, waters and glycerols permeate through the amphipathic pore of GlpF in a concerted, collective diffusion, driven or not driven by the osmotic pressure. If a deep enough chemical-potential well exists inside the channel (where the chemical potential is lower than the periplasm/cytoplasm bulk level), a glycerol molecule will be bound there, with a probability determined by the glycerol concentration and the dissociation constant. The bound glycerol will occlude permeation of waters and other glycerols through the channel. GlpF will switch between being open and closed to water permeation as a glycerol is dissociated from and bound to the binding site inside the protein's conducting pore. Therefore, such a chemical-potential profile means that glycerol modulates water permeation through GlpF in the glycerol concentration range around the dissociation constant.

In order to produce an accurate chemical-potential profile of glycerol, I conducted a total of 899 ns equilibrium and non-equilibrium molecular dynamics (MD) simulations, which amounts to about 10 times the computing efforts invested on GlpF in a published work of the current literature. The non-equilibrium MD simulations include three sets of steered molecular dynamics (SMD) [32-34] runs and two MD runs under pressure gradients. The accuracy of the chemical-potential estimation was ascertained by the agreement between the non-equilibrium SMD approach and the equilibrium adaptive biasing force (ABF) approach [24, 35].

Briefly, in this *in silico* study, glycerol is found to have a deep chemical-potential well in the GlpF channel near the Asn-Pro-Ala (NPA) motifs that is 6.5 kcal/mol below its chemical potential in the bulk of periplasm/cytoplasm, which corresponds to a dissociation constant of 14 µM. Glycerol binding to or dissociating from this binding site strongly modulates water permeation through the GlpF pore. There are two chemical-potential barriers separating the glycerol binding site from the periplasm and cytoplasm bulk regions; The barrier at the selectivity filter (SF) between the binding site and the periplasm is 10.1 kcal/mol, and the barrier between the NPA and the cytoplasm stands at 4.6 kcal/mol above the bottom of the chemical-potential well. This chemical-potential profile, considered in the context of the structural characteristics of GlpF, leads to a new theory of glycerol modulated water permeation through GlpF that is in agreement with the *in vitro* results in the current literature. It also harmonizes the existent theoretical results at 0 M glycerol concentration with the *in vitro* experiments at up to 100 mM concentrations of glycerol. Furthermore, it could be fully validated by future *in vitro* experiments measuring the glycerol-GlpF dissociation constant and the water permeability in the µM range of glycerol concentration.

**RESULTS AND DISCUSSION**

**Chemical-potential profile of glycerol.** Shown in Fig. 1 is the chemical potential of glycerol as a function of its center-of-mass position along a path leading from the periplasm to the entry vestibule of GlpF, through the channel, to the cytoplasm. Inside the channel, waters and glycerols line up in a single file, occluding one another from occupying the same z-coordinate. Outside the channel, farther away from the protein, there is more and more space for multiple waters and glycerols to occupy the same z-coordinate. In the single-file region, the chemical-potential is along the most probable path (minimal free-energy path, plotted in red in Fig. 1). In the non-single-file regions, the chemical potential curves (green) in Fig. 1 are along two straight lines leading from the periplasm to the channel entrance and from the channel exit to the cytoplasm. It needs to be noted here that chemical potential is not exactly identical to the potential of mean force (PMF) used in, *e.g.*, Refs. 25, 31. (See Eq. (2) in the next section.) In the non-single-file regions outside the conducting pore, there are many (infinite) possible paths for a glycerol to dissociate from GlpF. The PMF, being a function of only the z-coordinate of the glycerol's center of mass, necessarily involves the information of the glycerol and protein concentrations. In contrast, the chemical potential is computed here as a function of the glycerol's center-of-mass position (x-, y-, and z-coordinates) along two chosen straight lines leading from the protein to the periplasm and to the cytoplasm, respectively. It does not incorporate the concentration-dependence (namely, the entropic contributions) here. Considering this chemical potential (the standard free energy) along with the concentration terms[36, 37], one can show that the absolute binding free-energy of glycerol is equal to the chemical-potential difference between the apo state ($z = 30 \text{Å}$) and the bound state ($z = 6.5 \text{Å}$): $E_b = -6.5 \pm 1.5$ kcal/mol. Correspondingly, the dissociation constant of glycerol from GlpF is $k_d = \exp[E_b / k_B T] \sim 14\,\mu\text{M}$. It should be emphasized that the binding site is in the single-file region. Therefore, water permeation through GlpF is modulated by the glycerol concentration in the μM range because a glycerol bound inside GlpF occludes the conducting channel. The inhibitory concentration at half maximum (IC$_{50}$) is approximately 14 μM. Direct validation of this theory will require *in vitro* measurements of the dissociation constant of the glycerol-GlpF complex, which is currently unavailable in the literature. However, there are *in vitro* data available that validate the biophysical implications of this theory. Discussed below are the six implications of the chemical-potential profile given in Fig. 1.

*First, binding sites of glycerol in GlpF's pore.* The chemical-potential landscape (Fig. 1, top panel) corresponds well with the pore radius of GlpF along the channel (Fig. 1, middle and bottom panels). Where there is a maximum in pore radius, there is a minimum in chemical potential. There are three chemical-potential minima in the single-file region---one in the SF region and two in the NPA region. The locations of these minima are in agreement of the structural studies of GlpF that show a glycerol at the SF and another at the NPA [17, 18]. The two minima near the NPA are both far below the bulk chemical-potential level ($< -5$ kcal/mol). They are separated from one another by a very low barrier (<3 kcal/mol) and, therefore, a glycerol bound inside the GlpF pore can traverse the region of $2\text{Å} < z < 7\text{Å}$ rather freely. The minimum at the SF ($z = -4\text{Å}$) is above the bulk chemical-potential level and it is rather shallow. The barrier separating this minimum from the periplasm is only 2.5 kcal/mol. Dissociation of a glycerol bound at the SF is expected to be within nanoseconds and, indeed, it was observed so in the equilibrium MD simulation. It should be noted that Ref. 21 already gave an PMF curve with minima at the SF and NPA that correspond to the binding sites of glycerol determined in the structural study of Ref. 17. However, in terms of glycerol modulating water permeation, the IC$_{50}$ based on their PMF curve would be 24 mM, which is inconsistent with the *in vitro* study that the presence of up to 100 mM glycerol did not affect the kinetics of water transport[13].

*Second, activation barrier for glycerol permeation.* The glycerol permeation rate was measured at multiple temperatures that gave an Arrhenius activation barrier of 9.6±1.5 kcal/mol [13]. This corresponds precisely to the chemical-potential barrier at the SF that is 10.1±1.2 kcal/mol above the bottom of the chemical-potential well near the NPA. It should be noted that the agreement between this theoretical result and the *in vitro* data comes directly, without any adjustments, from the *in silico* experiments based on the all-atom CHARMM36 force field [38]. Also using the all-atom CHARMM force field, Jensen *et al*[21] and Henin *et al*[24] obtained similar values of the activation barrier for glycerol permeation. However, their overall free-energy profiles differ significantly from one another and from that of this work (Fig. 1) obtained with more than ten times the computing efforts in either of them. It should also be noted that, to compute the glycerol permeation rate from the chemical-potential profile, one needs to convert the (x,y,z)-dependent chemical potential into the z-dependent PMF, including the entropic penalty term. [21, 25, 31] (See Eq.(2) in the next section.) The entropic penalty is proposal to the absolute temperature and, therefore, does not alter the Arrhenius activation barrier but modifies the pre-factor of the Arrhenius rate formula.

*Third, activation barrier for water permeation.* In the bound state, a glycerol molecule resides in the single-file region, deep inside GlpF's conducting channel, occluding waters or other glycerols from traversing the channel. When the glycerol concentration is in the mM range (far above the dissociation constant, 14 µM), the GlpF pore is practically saturated with glycerol. Water permeation is fully correlated with the glycerol dissociation. Therefore, the Arrhenius barrier for water permeation through GlpF at mM glycerol concentrations is no less than the activation barrier for glycerol dissociation to the cytoplasm (6.5 kcal/mol*)*. This theoretical result of the activation barrier is in agreement with the *in vitro* data of 7 kcal/mol derived from measurements of the water permeation rates at multiple temperatures [13] if one assumes that the glycerol concentration in the *in vitro* experiments was significantly higher than the $IC_{50}$ of 14 µM. Indeed, 10% (volume/volume) glycerol was used in the experiments throughout the protein preparation procedures of Ref. 13 even in the wash buffer to remove nonspecifically bound material from GlpF. The proteins so prepared were saturated with glycerols that were not removed from the protein. Those glycerols were brought into the transport assay buffer but they were not counted in the glycerol concentration of the buffer. Therefore, the 0 mM end of their glycerol concentration range (up to 100 mM) can be less than 1 mM but still much higher than 14 µM. At 0 µM glycerol concentration, GlpF's conducting pore will be glycerol-free. The chemical-potential of water permeation was determined to be flat throughout the open channel, having a low Arrhenius barrier, $E_a^0 < 4\,\text{kcal/mol}$ [20, 27, 31], similar to that of AQPZ [14].

*Fourth, water permeation rates through glycerol-free vs. glycerol-saturated GlpF. In vitro* experiments demonstrated that the water permeation rate in the range of glycerol concentration up to 100 mM was much lower than the water-specific aquaporins (including AQPZ) [13, 14, 30], but theoretical studies showed that the rate of water permeation through GlpF in absence of glycerols is similar to or greater than through AQPZ [23, 31]. This puzzling point actually serves to validate our theory: Water permeation through glycerol-bound GlpF is much slower than through apo GlpF because a glycerol bound inside GlpF's conducting channel occludes waters and the dissociation of this bound glycerol is a much slower process than that of a water traversing the glycerol-free channel of GlpF. However, as pointed out in Refs. [23, 31], the permeation rate measurements are directly related to the reconstitution efficiencies in experiments that are often difficult to control [14]. In contrast, the Arrhenius activation barrier is independent of the reconstitution efficiency. Agreement between its theoretical value and the *in vitro* results gives a higher degree of certainty than comparing the permeation rate values. Nevertheless, I conducted two equilibrium MD simulations of 100 ns each to compute the rates of water permeation through GlpF with 0 M and 14 mM glycerol concentrations, respectively. The osmotic permeability of water at 0 M glycerol concentration was computed to be $(11\pm 1)\times 10^{-14}\,cm^3/s$, which is in agreement with other MD studies in the current literature [23]. And the osmotic permeability at 14 mM glycerol concentration was found to be $(1.5\pm 0.2)\times 10^{-14}\,cm^3/s$, which is consistent with the *in vitro* findings [16], assuming that the glycerol concentration "0M" in their experiments was actually much higher than the IC$_{50}$ of 14 µM. This assumption is justified by their use of 10% (volume/volume) glycerol throughout the protein preparation procedures during which GlpF were saturated with glycerols. Those glycerols were not removed from the protein but brought into the experiment of transport assays. The count of glycerols in transport assay buffers did not include the glycerols brought in by the proteins and, therefore, the glycerol concentration of "0M" was not exactly 0 mM but non-zero sub-mM.

*Fifth, concentration-dependence.* There is currently no *in vitro* data of the water permeation rate in the µM range of glycerol concentration. Functional experiments in this concentration range are expected to demonstrate significant variation in the water permeation rate from the glycerol-free limit to the glycerol-saturated limit.

*Sixth, direction-dependence of water permeation.* The glycerol chemical-potential landscape is asymmetrical between the cytoplasmic and the periplasmic sides. This does not mean that there is asymmetry in glycerol transport [22] but it may imply that high glycerol concentrations can make water permeation through GlpF direction-dependent: Permeation from the periplasm to the cytoplasm has an Arrhenius activation barrier of 6.5 kcal/mol while permeation from the cytoplasm to the periplasm has a barrier of 10.1 kcal/mol. A glycerol molecule bound inside the GlpF practically occludes water permeation from the cytoplasm to the periplasm, making water permeation through saturated GlpF directional. In order to explicitly test this possibility, I conducted two nonequilibrium MD simulations of SysI in which the waters in and within the vicinity of the conducting pore are subject to a pressure gradient. Under a pressure-induced force field of $0.01 \, \text{kcal}/\text{mol} \cdot \text{Å} \cdot \text{amu}$ in the positive z-direction, water permeation flow was observed at 50 events/ns, on the average, through each channel. Under the force field of the same magnitude but in the negative z-direction, no water permeation events were observed within 4 ns. It should be pointed out that these two nonequilibrium MD simulations cannot be taken as quantitatively accurate because the applied pressure is many orders of magnitude above a realistic osmotic pressure in a biological system. However, they may correctly point to the asymmetry of water permeation at high glycerol concentrations. Another point worth noting is that an imbalance in glycerol concentrations between the periplasm and the cytoplasm may have measurable effects on the water permeation through GlpF as well.

**Relevant interactions.** Now, what interactions are responsible for the chemical-potential profile shown in Fig. 1? Full understanding has been achieved in regards to the hydrogen-bonding of glycerol/water to the luminal atoms of GlpF throughout its amphipathic pore [17, 20, 21]. These interactions do not give rise to high barriers or deep wells because the number of hydrogen bonds that a glycerol/water can form with GlpF does not vary drastically with its location throughout the conducting channel. However, the dimension of the conducting pore is not uniform (Fig. 1, middle panel). The narrowest part of the channel is in the SF region, and another narrow part is between the NPA motifs and the channel exit on the cytoplasmic side. In the region around the NPA motifs, the channel is still single-file in nature but wider than the two narrow regions. Interestingly enough, even the narrowest part of the GlpF channel ($3\text{Å}$ in diameter) is wider than the diameter of a water molecule ($2\text{Å}$). Water can traverse the entire channel without causing structural distortions to the protein. Permeation of water, in the absence of glycerols, is dominated by the hydrogen-bonding interactions among waters and with the luminal atoms. In contrast, the size of a glycerol molecule is $3.1\text{Å}$ in the least extended dimension. It cannot traverse the GlpF pore as freely as a water molecule can. In fact, it is the close contact between a glycerol molecule and the GlpF lumen that gives rise to the chemical-potential profile shown in Fig. 1. The width of the pore turns out to be such that the van der Waals (VDW) forces on the glycerol by the surrounding residues are all attractive in the NPA region, causing the chemical potential there to be lower than the bulk level.

Fig. 2(A) shows that, when a glycerol is in the SF region ($z = -3.1\text{Å}$), the narrowest part of the GlpF channel, four residues (Trp 48, Gly 199, Phe 200, and Arg 206) are within the repulsive range (less than $2\text{Å}$) of the glycerol. The VDW interactions between these four residues and the glycerol are repulsive, leading to the peak of the VDW energy around $z = -3\text{Å}$ shown in Fig. 2(D).

Fig. 2(B) shows that, when a glycerol is near the NPA motifs ($z = 4.0\text{Å}$), the widest part of the GlpF channel, there are eight residues (Leu 21, Val 52, Leu 67, Asn 68, Leu 159, Ile 187, Met 202, and Asn 203) in the attractive range (between $2\text{Å}$ and $3\text{Å}$) of the glycerol. The attractive VDW interactions between these residues and the glycerol are responsible for the minimum of the VDW energy in the NPA region shown in Fig. 2(D), which ultimately leads to the chemical-potential well in Fig. 1.

Fig. 2(C) shows that, when a glycerol is located in between the NPA and the pore exit ($z = 10 Å$), the other narrow part of the GlpF channel, there are three residues (Ala 65, His 66, and Leu 67) in the repulsive range (less than $2 Å$) of the glycerol. The repulsive VDW interactions between these three residues and the glycerol are weaker than when the glycerol is located near the SF. They show up in Fig. 2(D) as a barrier albeit lower than the one at the SF, and they are ultimately responsible for the barrier in the chemical-potential profile on the cytoplasmic side of the NPA (Fig. 1).

Fig. 2(E) shows the dihedral energy of glycerol as a function of its center-of-mass z-coordinate. Figs. 2 (F) and (G) show the root-mean-square deviations of the afore-listed residues from their apo, equilibrium position as functions of the glycerol's center-of-mass z-coordinate. Among the three groups of residues, only the SF-forming residues (Fig. 2(A)) deviate significantly from their apo, equilibrium positions when the glycerol passes by there.

All the factors shown in Fig. 2 combine to demonstrate that the bound state of a glycerol deep inside the GlpF channel owes its existence to the attractive VDW interactions between glycerol and the residues near the NPA motifs. The two chemical-potential barriers result mainly from the repulsive VDW forces on the glycerol by the residues in the two narrow parts of the GlpF channel. The barrier at the SF also has significant contributions from the conformational changes of the SF-forming residues caused by the presence of the glycerol and from the conformational changes of the glycerol itself.

**METHODS**

**System setup.** This study was based on the following all-atom model of GlpF in the cell membrane (Fig. 3): The GlpF tetramer, formed from the crystal structure (PDB code: 1FX8) with 12 glycerols, was embedded in a patch of fully hydrated palmitoyloleylphosphatidyl-ethanolamine (POPE) bilayer. The GlpF-POPE complex is sandwiched by two layers of water, each of which is approximately $40 Å$ in thickness. The system is neutralized and ionized with Na$^+$ and Cl$^-$ ions at a concentration of 111 mM. The entire system, consisting of 150,855 atoms, is 114Å×115 Å×112 Å in dimension when fully equilibrated. This system (SysI) has a glycerol concentration of 14 mM. A second system (SysII), for 0 M glycerol, was derived from SysI by deleting all 12 glycerols. It has 150,687 atoms in all and dimensions approximately equal to those of SysI. The Cartesian coordinates are chosen such that the origin is at the geometric center of the GlpF tetramer. The xy-plane is parallel to the lipid-water interface and the z–axis is pointing from the periplasm to the cytoplasm.

All the simulations of this work were performed using NAMD 2.8 [39]. The all-atom CHARMM36 parameters [38, 40] were adopted for all the inter- and intra-molecular interactions. Water was represented explicitly with the TIP3 model. The pressure and the temperature were maintained at 1 bar and 293.15 K, respectively. The Langevin damping coefficient was chosen to be 5/ps. The periodic boundary conditions were applied to all three dimensions, and the particle mesh Ewald was used for the long-range electrostatic interactions. Covalent bonds of hydrogens were fixed to their equilibrium length. The time step of 2 fs was used for short-range interactions in equilibrium simulations, but 1 fs was used for nonequilibrium runs. The same time step of 4 fs was used for long-range forces in both equilibrium and nonequilibrium simulations. The cut-off for long-range interactions was set to 12Å with a switching distance of 10Å. In all simulations, the alpha carbons on the trans-membrane helices of GlpF within the range of −10Å<$z$<10Å were fixed to fully respect the crystal structure.

**Equilibrium MD.** Two runs of 100 ns each in length were conducted for SysI and SysII respectively. Using the theoretical formulation of Ref. 41, I computed the mean square displacements (MSD) of the water molecules in the conducting pore. (The MSD curves are shown in supplemental Fig. S1.) The slope of the MSD curve gives an estimate of the osmotic permeabilities of both SysI and SysII that were presented in the previous section.

**Non-equilibrium SMD.** SMD runs were conducted to sample the transition paths of glycerol going from the periplasm, through the conducting pore, to the cytoplasm, for computing the chemical-potential of glycerol as a function of its center-of-mass position. Three sets of SMD simulations were completed to achieve reliable statistics. (The details are given in the supplemental Figs. S2 to S4.) In each set of SMD, the starting structure is the fully equilibrated structure of SysI. The center of mass of a glycerol was steered/pulled in the positive z-direction to sample a forward pulling path, and then pulled in the negative z-direction to sample a reverse pulling path. In Set #1 and Set #2, the entire path leading from the periplasm, through the GlpF pore, to the cytoplasm bulk region, was divided into four sections: one section in the periplasm, one section in the cytoplasm, and two sections in the single-file region of the conducting pore. In Set #3, the path through the conducting pore was divided into 19 sections, each 1Å in width. Four forward paths and four reverse paths were sampled in each section of Set #1 and the same were done in Set #2. Ten forward paths and ten reverse paths were sampled in each section of Set #3. In all three sets, 4 ns equilibration was performed at both end points of each section so that the pulling paths were sampled between equilibrium states with the glycerol's center-of-mass coordinates being fixed at desired values. And, in all cases, the pulling speed was v=2.5 Å /ns.

Along each forward pulling path from A to B, the work done to system was recorded as $W_{A \to Z}$ when glycerol was pulled from A to Z. Along each reverse pulling path from B to A, the work done to system was recorded as $W_{B \to Z}$ when glycerol was pulled from B to Z. Here Z represents a state of the system when the center-of-mass z-coordinate of the pulled glycerol is z. A and B represent two end states of a given section, respectively. The chemical potential (the standard free energy[36]) of glycerol, $\mu(x, y, z)$, a function of its center-of-mass position along the permeation path, can be computed from the Brownian dynamics fluctuation-dissipation theorem (BD-FDT) as follows [27, 34]:

$$\mu(x, y, z) - \mu(x_A, y_A, z_A) = -k_B T \ln \left( \frac{\langle \exp[-W_{A \to Z} / 2k_B T] \rangle_F}{\langle \exp[-W_{Z \to A} / 2k_B T] \rangle_R} \right). \tag{1}$$

Here $W_{Z \to A} = W_{B \to A} - W_{B \to Z}$ is the work done to the system for the part of a reverse path when the glycerol was pulled from Z to A. $k_B$ is the Boltzmann constant and $T$ is the absolute temperature. $z_A$ and $z_B$ are the z-coordinates of the center of mass of the pulled glycerol at the end states A and B of the system, respectively. From the chemical potential, one can approximate the potential of mean force as a function of only the z-coordinate:

$$G(z) = -k_B T \ln\left( \int dxdy \exp\left[-\mu(x, y, z)/k_B T\right] / A_{\text{ref.}} \right)$$
$$\simeq \mu(x^*(z), y^*(z), z) - k_B T \ln(A(z)/A_{\text{ref.}})$$
(2)

where $A_{\text{ref}}$ and $A(z)$ are, respectively, the area for reference and the area occupied by the center of mass of glycerol on the plane of a given z-coordinate. They are directly related to the glycerol and protein concentrations. $x^*(z)$ and $y^*(z)$ are the median coordinates of integration on the same plane. The second term of the second line of Eq. (2) is the entropic penalty. This term is proportional to the temperature. It does modify the overall rate of transport but does not alter the Arrhenius activation barrier.

**Accuracy confirmed with ABF approach.** In order to ascertain the accuracy of the chemical-potential profile determined in the BD-FDT approach, I applied the mature ABF method [24, 35] to two 1Å-wide windows ($-5.5\text{Å} < z < -4.5\text{Å}$ and $-4.5\text{Å} < z < -3.5\text{Å}$). A 20 ns MD run and a 90 ns MD run were carried out for the two windows respectively so that convergence was achieved in each window. (The results are shown in the supplemental Figs. S5 and S6 respectively.) They confirm the accuracy of the chemical-potential profile in Fig. 1. Note that the ABF method, an equilibrium sampling approach, is distinctive in nature from the BD-FDT, a non-equilibrium sampling method. Yet, the two approaches produced identical results when long enough MD runs were conducted.

**Non-equilibrium MD under pressure gradients.** All the conditions were identical to the equilibrium MD runs of SysI except that the waters (not glycerols) within the conducting pore were subject to a constant force acting on their centers of mass. Those waters have their coordinates within the following range:

$$8\text{Å} <|x|< 20\text{Å}, 8\text{Å} <|y|< 20\text{Å}, |z|<10\text{Å} \tag{3}$$

Starting from the end state of the equilibrium MD run of SysI, a 4 ns run was conducted under the afore-described force in the z-direction and the same was done for the negative z-direction.

**CONCLUSION**

Based on the existent *in vitro* experiments and the chemical-potential profile mapped out in the present study, it is logical to suggest that glycerol bound inside GlpF's conducting channel modulates water permeation through its amphipathic pore. This mechanism explains why GlpF's water-permeability is significantly lower than the water-specific aquaporins and why water permeation has a large Arrhenius activation barrier when the glycerol concentration is in the mM range. As an addition to the science of hydrogen-bonding of waters and glycerols in the conducting pore, this study demonstrates that the van der Waals interactions between the GlpF and a glycerol play a distinctive biological role. The size of the conducting pore is such that a region exists near the NPA motifs where the VDW interactions between the GlpF and a glycerol are attractive. This precise steric arrangement of GlpF causes the glycerol's chemical potential there to be lower than its bulk level and, therefore, a bound state of glycerol exists deep inside the single-file channel. Finally, the conclusions of this study of GlpF may be applicable to other members of the aquaglyceroporin sub-family of the major intrinsic proteins.

**ACKNOWLEDGEMENTS**

The author acknowledges support from the NIH (Grant #GM084834) and the Texas Advanced Computing Center.

Bibliographic references and notes

1.  P. Agre, M. Bonhivers and M. J. Borgnia, *J Biol Chem*, 1998, **273**, 14659-14662).
2.  M. Borgnia, S. Nielsen, A. Engel and P. Agre, *Annu Rev Biochem*, 1999, **68**, 425-458).
3.  J. B. Heymann and A. Engel, *Physiology*, 1999, **14**, 187-193).
4.  J. K. Lee, S. Khademi, W. Harries, D. Savage, L. Miercke and R. M. Stroud, *Journal of Synchrotron Radiation*, 2004, **11**, 86-88).
5.  P. Agre, *Neurosci. Res.*, 2008, **61**, S2-S2).
6.  J. M. Carbrey and P. Agre, ed. E. Beitz, Springer Berlin Heidelberg, 2009, vol. 190, pp. 3-28.
7.  K. B. L. Heller, E. C. C.; Wilson, T. Hastings *Journal of Bacteriology*, 1980, **144**, 5).
8.  M. L. Zeidel, S. V. Ambudkar, B. L. Smith and P. Agre, *Biochemistry*, 1992, **31**, 7436-7440).
9.  C. Maurel, J. Reizer, J. I. Schroeder, M. J. Chrispeels and M. H. Saier, *Journal of Biological Chemistry*, 1994, **269**, 11869-11872).
10. T. Walz, B. L. Smith, M. L. Zeidel, A. Engel and P. Agre, *Journal of Biological Chemistry*, 1994, **269**, 1583-1586).
11. M. L. Zeidel, S. Nielsen, B. L. Smith, S. V. Ambudkar, A. B. Maunsbach and P. Agre, *Biochemistry*, 1994, **33**, 1606-1615).
12. B. Yang and A. S. Verkman, *Journal of Biological Chemistry*, 1997, **272**, 16140-16146).
13. M. J. Borgnia and P. Agre, *Proc Natl Acad Sci U S A*, 2001, **98**, 2888-2893).
14. P. Pohl, S. M. Saparov, M. J. Borgnia and P. Agre, *Proc Natl Acad Sci U S A*, 2001, **98**, 9624-9629).
15. M. Hansen, J. F. J. Kun, J. E. Schultz and E. Beitz, *Journal of Biological Chemistry*, 2002, **277**, 4874-4882).
16. S. M. Saparov, S. P. Tsunoda and P. Pohl, *Biol. Cell*, 2005, **97**, 545-550).
17. D. Fu, A. Libson, L. J. W. Miercke, C. Weitzman, P. Nollert, J. Krucinski and R. M. Stroud, *Science*, 2000, **290**, 481-486).
18. E. Tajkhorshid, P. Nollert, M. Ø. Jensen, L. J. W. Miercke, J. O'Connell, R. M. Stroud and K. Schulten, *Science*, 2002, **296**, 525-530).
19. J. Song, A. Almasalmeh, D. Krenc and E. Beitz, *Biochimica et Biophysica Acta (BBA) - Biomembranes*, 2012, **1818**, 1218-1224. (Results (Fig. 1, in particular): In presence of sorbitol, expressions of Glpf or PfAQP in protoplasts did not increase water permeabilities. In presence of sucrose, expression of Glpf in protoplasts did not increase water permeability but expression of PfAQP did increase water permeability slightly. Conclusions one can draw: sucrose inhibits GlpF more strongly than PfAQP. But no conclusion as to whether or not glycerol modulates water permeation through GlpF).
20. B. L. de Groot and H. Grubmüller, *Science*, 2001, **294**, 2353-2357).
21. M. Ø. Jensen, S. Park, E. Tajkhorshid and K. Schulten, *Proceedings of the National Academy of Sciences*, 2002, **99**, 6731-6736).
22. D. Lu, P. Grayson and K. Schulten, *Biophysical journal*, 2003, **85**, 2977-2987).
23. M. Ø. Jensen and O. G. Mouritsen, *Biophysical journal*, 2006, **90**, 2270-2284).
24. J. Hénin, E. Tajkhorshid, K. Schulten and C. Chipot, *Biophysical journal*, 2008, **94**, 832-839).
25. J. S. Hub and B. L. de Groot, *Proc Natl Acad Sci U S A*, 2008, **105**, 1198-1203).
26. L. Y. Chen, *Biophysical Chemistry*, 2010, **151**, 178-180).
27. L. Y. Chen, D. A. Bastien and H. E. Espejel, *Physical Chemistry Chemical Physics*, 2010, **12**, 6579-6582).
28. R. Oliva, G. Calamita, J. M. Thornton and M. Pellegrini-Calace, *Proceedings of the National Academy of Sciences*, 2010, **107**, 4135-4140).
29. N. Chakrabarti, B. Roux and R. Pomès, *Journal of molecular biology*, 2004, **343**, 493-510).

30. M. J. Borgnia, D. Kozono, G. Calamita, P. C. Maloney and P. Agre, *Journal of molecular biology*, 1999, **291**, 1169-1179).
31. C. Aponte-Santamaria, J. S. Hub and B. L. de Groot, *Phys Chem Chem Phys*, 2010, **12**, 10246-10254).
32. B. Isralewitz, J. Baudry, J. Gullingsrud, D. Kosztin and K. Schulten, *Journal of Molecular Graphics and Modelling*, 2001, **19**, 13-25).
33. S. Park and K. Schulten, *The Journal of Chemical Physics*, 2004, **120**, 5946-5961).
34. L. Y. Chen, *Physical Chemistry Chemical Physics*, 2011, **13**, 6176-6183).
35. E. Darve and A. Pohorille, *The Journal of Chemical Physics*, 2001, **115**, 9169-9183).
36. I. N. Serdyuk, Zaccai, Nathan R., and Zaccai, Joseph, *Methods in Molecular Biophysics Structure, Dynamics, Function*, Cambridge University Press, Cambridge, 2007. Section C1.3.
37. D. Chandler, *Introduction to Modern Statistical Mechanics*, Oxford University Press, Oxford, 1987. Section 7.3.
38. A. D. MacKerell, N. Banavali and N. Foloppe, *Biopolymers*, 2000, **56**, 257-265).
39. J. C. Phillips, R. Braun, W. Wang, J. Gumbart, E. Tajkhorshid, E. Villa, C. Chipot, R. D. Skeel, L. Kalé and K. Schulten, *Journal of Computational Chemistry*, 2005, **26**, 1781-1802).
40. B. R. Brooks, C. L. Brooks, A. D. Mackerell, L. Nilsson, R. J. Petrella, B. Roux, Y. Won, G. Archontis, C. Bartels, S. Boresch, A. Caflisch, L. Caves, Q. Cui, A. R. Dinner, M. Feig, S. Fischer, J. Gao, M. Hodoscek, W. Im, K. Kuczera, T. Lazaridis, J. Ma, V. Ovchinnikov, E. Paci, R. W. Pastor, C. B. Post, J. Z. Pu, M. Schaefer, B. Tidor, R. M. Venable, H. L. Woodcock, X. Wu, W. Yang, D. M. York and M. Karplus, *Journal of Computational Chemistry*, 2009, **30**, 1545-1614).
41. F. Zhu, E. Tajkhorshid and K. Schulten, *Phys Rev Lett*, 2004, **93**, 224501).
42. W. Humphrey, A. Dalke and K. Schulten, *Journal of Molecular Graphics*, 1996, **14**, 33-38).
43. O. S. Smart, J. M. Goodfellow and B. A. Wallace, *Biophysical journal*, 1993, **65**, 2455-2460).

**Figures**

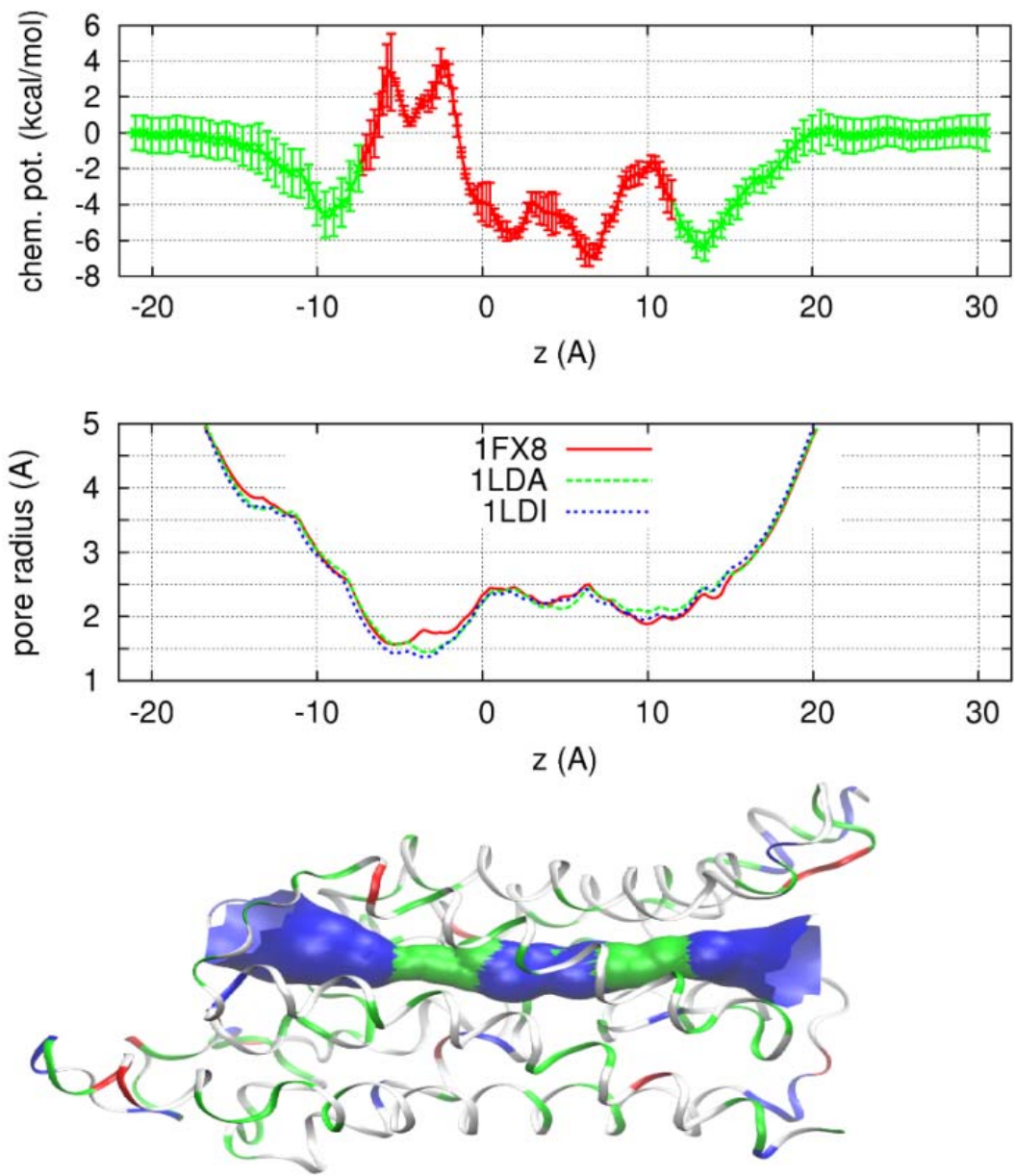

Fig. 1. The chemical potential of glycerol as a function of its center-of-mass z-coordinate (top panel) where the error bars indicate standard deviation (SD). The pore radius of GlpF crystals (middle panel) from the RCSB Protein Data Bank whose access codes are as shown. The SF region is around $z = -3 Å$ and the NPA motifs are located around $z = 4 Å$. The bottom panel illustrates the protein structure (in ribbons, colored with residue type) and the conducting pore of GlpF. Protein structure rendered with VMD [42] and pore radius with HOLE2 [43].

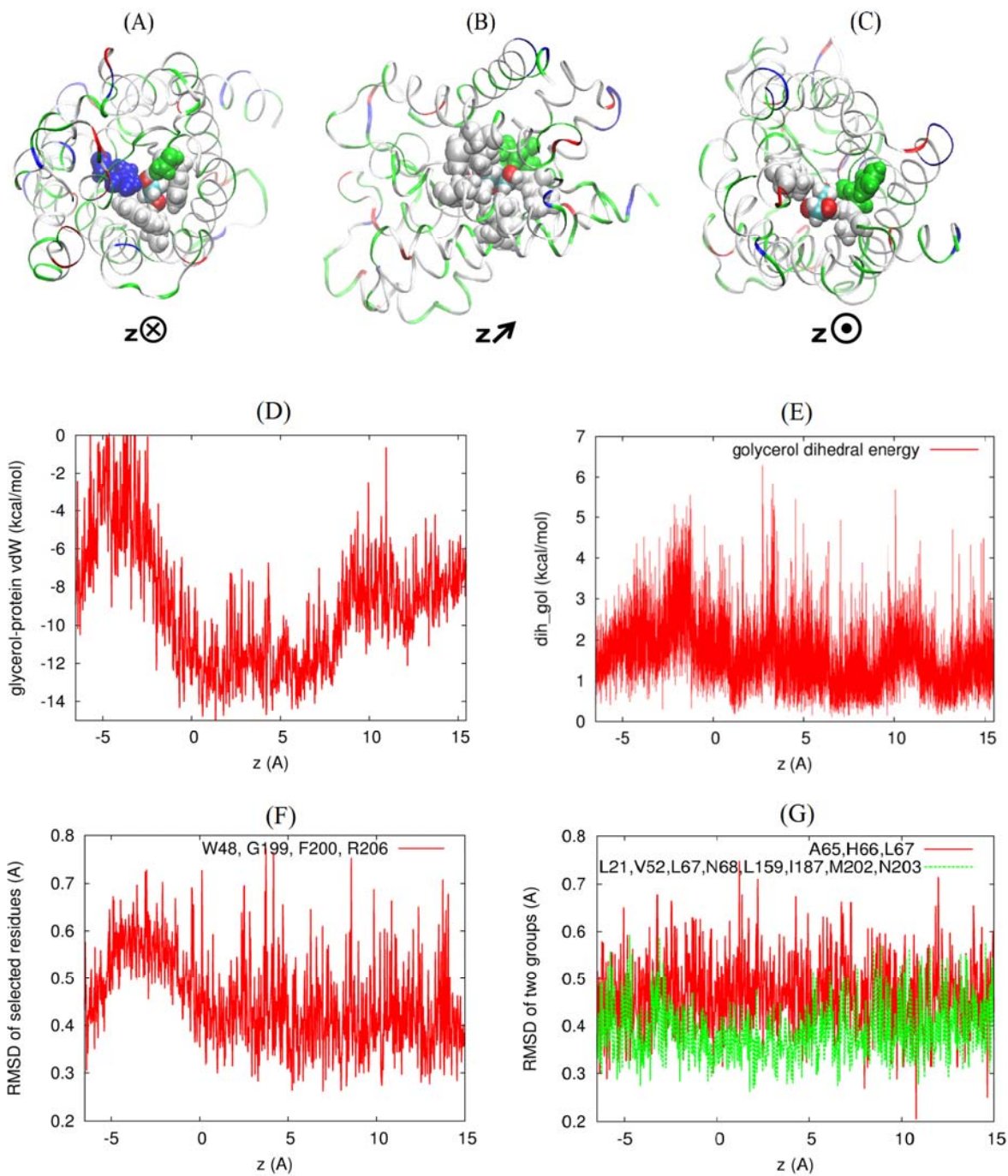

Fig. 2. The glycerol-GlpF VDW interactions. Top panel, a protomer (in ribbons, colored with residue type) and the glycerol (in vdw, colored with atom name) whose center of mass is located at (A) $z = -3.1 \text{Å}$, (B) $z = 4.0 \text{Å}$, and (C) $z = 10 \text{Å}$ respectively. The orientations of the protein are illustrated with the direction of the z-axis pointing (A) into the plane, (B) slanted, and (C) out of the plane. Also shown in the top panel are the relevant residues (in vdw, colored with residue type): (A) Trp 48, Gly 199, Phe 200, and Arg 206 that are in the repulsive range (less than $2 \text{Å}$) of glycerol, (B) Leu 21, Val 52, Leu 67, Asn 68, Leu 159, Ile 187, Met 202, and Asn 203 that are in the attractive range (between $2 \text{Å}$ and $3 \text{Å}$) of glycerol, and (C) Ala 65, His 66, and Leu 67 that are in the repulsive range (less than $2\text{Å}$) of glycerol. (D) The VDW interaction between a glycerol molecule and GlpF throughout the permeation pore. (E) The di-hedral energy of glycerol as a function of its center-of-mass z-coordinate. (F) The RMSD of selected residues shown in (A) as a function of the glycerol's center-of-mass z-coordinate. (G) The same for the residues shown in (B) and (C), respectively. Protein structures rendered with VMD [42].

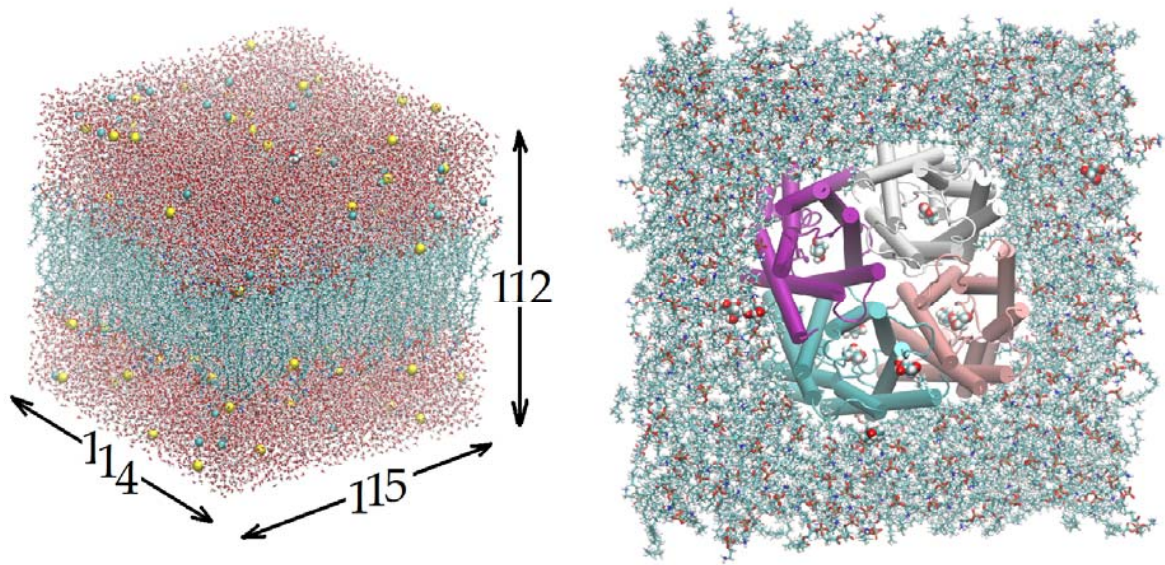

Fig. 3. All-atom model of GlpF in the cell membrane. The system is $114\text{Å} \times 115\text{Å} \times 112\text{Å}$ in dimension. Visible in the left panel are waters (in licorice representation), lipids (licorice), ions (vdw), and one glycerol (vdw). Shown in the right panel are the GlpF tetramer (in cartoon representation, colored by segname), lipids (licorice), and glycerols (vdw). All except GlpF are colored by element name. Graphics rendered with VMD [42].

Supplementary figures in support of "Glycerol modulates water permeation through *Escherichia coli* aquaglyceroporin GlpF"

L. Y. Chen

University of Texas at San Antonio

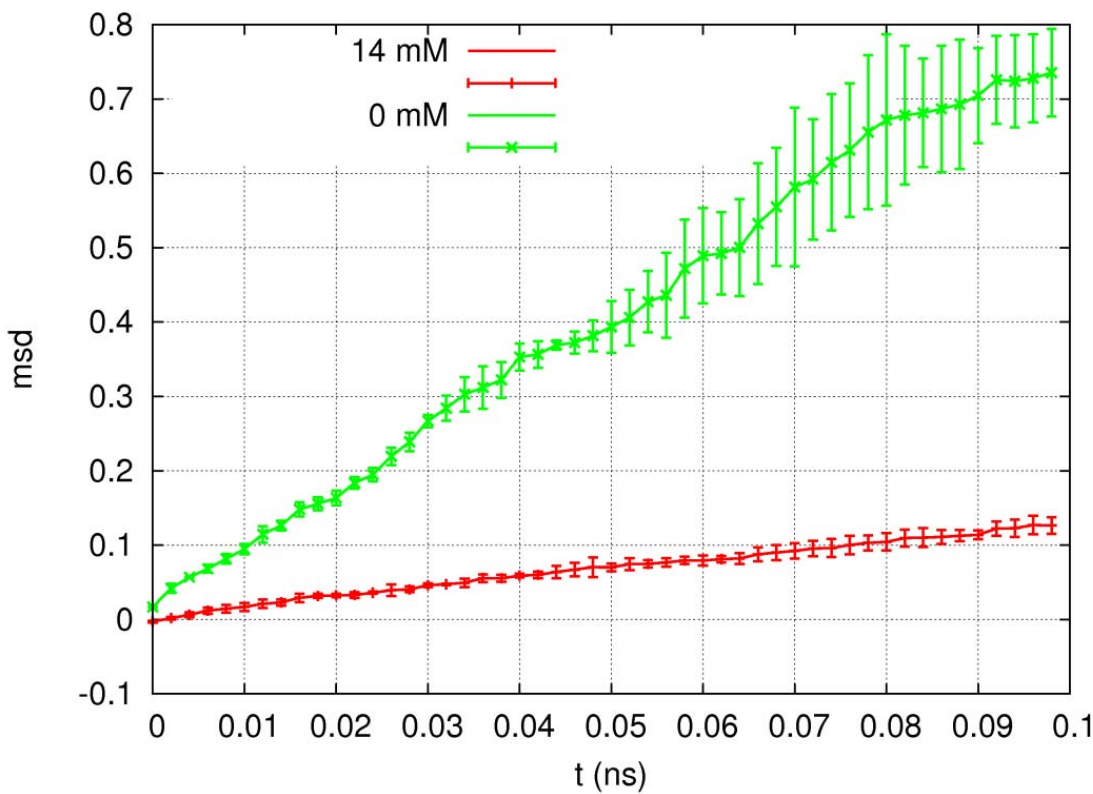

Fig. S1. The mean square displacement (msd) of waters in the conducting pore of GlpF, $\langle \delta n^2(t) \rangle$, as defined in Refs. (1, 2). The mean and standard deviation were taken over three sets of 10 ns trajectories (70 to 80 ns, 80 to 90 ns, and 90 to 100 ns) from the 100 ns equilibrium MD run for each system. In each set, the 10 ns MD trajectory was divided into 100 segments of 0.1 ns each, over which the displacements of the waters in the pore were included in the relevant statistics. The osmotic permeation rate of water, $p_f = v_W D_n$, is related to the slope of an msd curve, $D_n = \langle \delta n^2(t) \rangle / 2t$ (1, 2). Here $v_W = 3 \times 10^{-23} \, cm^3$ is the average volume occupied by one water molecule.

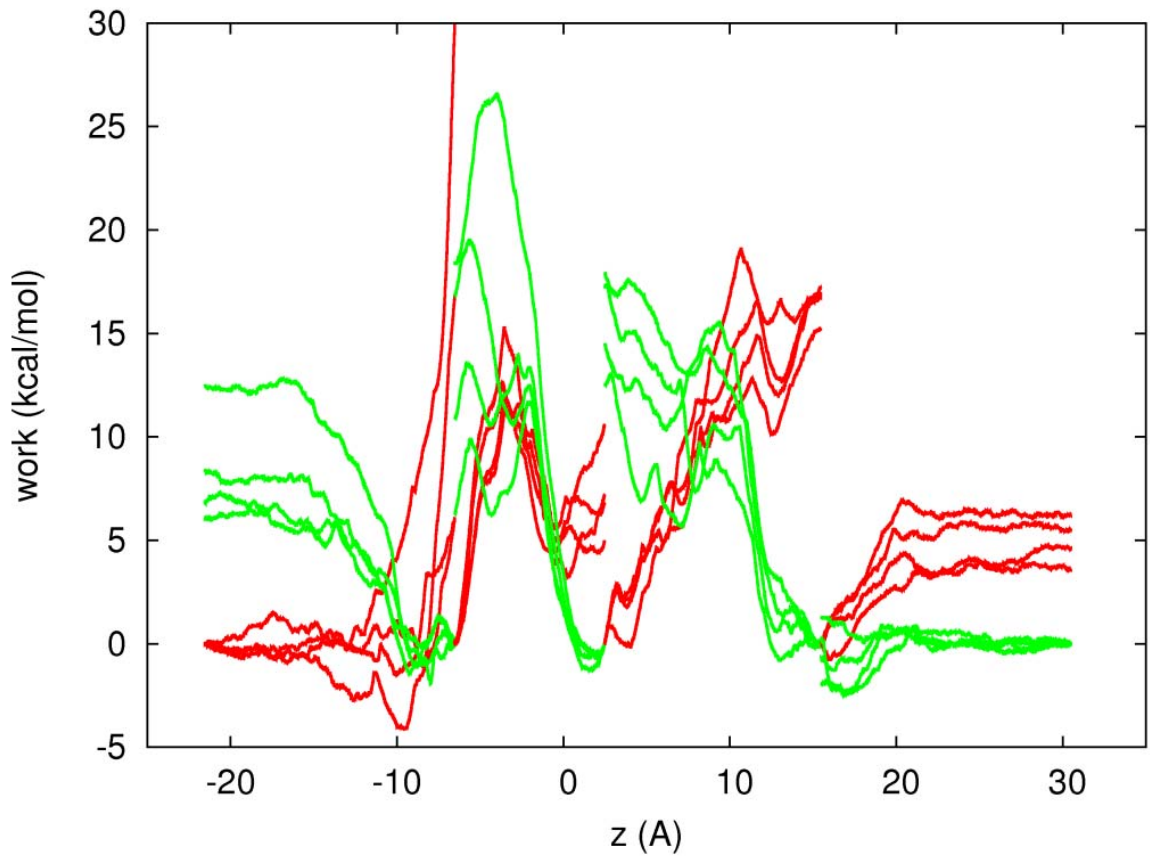

Fig. S2. Work done to the system as a function of the glycerol's center-of-mass z-coordinate along the forward (red) and reverse (green) pulling paths in SMD simulation Set #1. From the periplasm to the entry vestibule of GlpF, $-21.5\text{Å} < z < -6.5\text{Å}$, the glycerol center-of-mass was pulled along a straight line. Through the single-file channel of GlpF, $-6.5\text{Å} < z < 15.5\text{Å}$, the glycerol center-of-mass was pulled along the z-direction while its x- and y-degrees of freedom were not restricted. From the GlpF exit to the cytoplasm, $15.5\text{Å} < z < 30.5\text{Å}$, the pulling was along a straight line. The pulling speed was $2.5\text{Å}/ns$. Long time equilibrations (4 ns each) were done at every end of each segment, $z = -21.5\text{Å}, -6.5\text{Å}, 4.5\text{Å}, 14.5\text{Å},$ and $30.5\text{Å}$, respectively, so that the SMD pulling was between equilibrium states at the two ends of each segment.

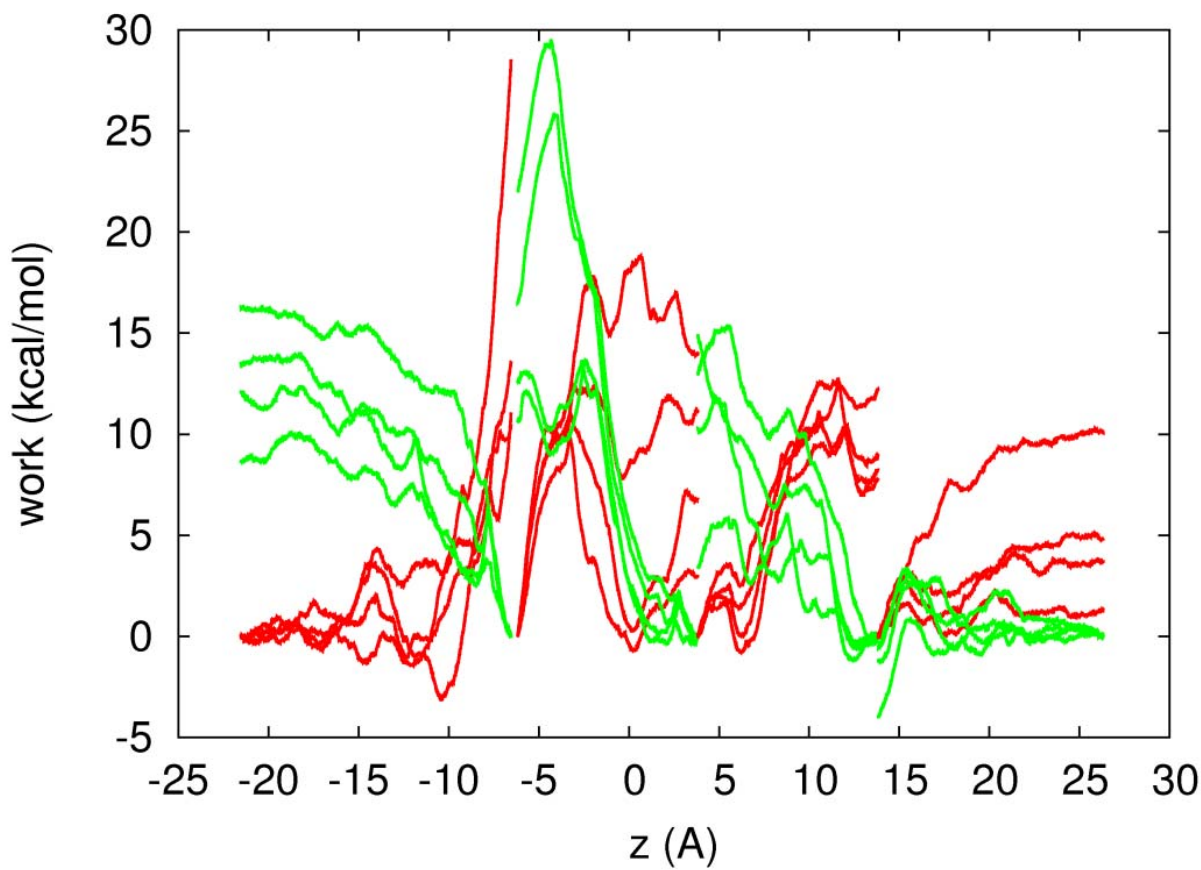

Fig. S3. Work done to the system as a function of the glycerol's center-of-mass z-coordinate along the forward (red) and reverse (green) pulling paths in SMD simulation Set #2. From the periplasm to the entry vestibule of GlpF, $-21.5\text{Å} < z < -6.5\text{Å}$, the glycerol center-of-mass was pulled along a straight line. Through the single-file channel of GlpF, $-6.5\text{Å} < z < 14.5\text{Å}$, the glycerol center-of-mass was pulled along the z-direction while its x- and y-degrees of freedom were not restricted. From the GlpF exit to the cytoplasm, $14.5\text{Å} < z < 26.5\text{Å}$, the pulling was along a straight line. The pulling speed was $2.5\text{Å}/ns$. Long time equilibrations (4 ns each) were done at every end of each segment, $z = -21.5\text{Å}, -6.5\text{Å}, 4.5\text{Å}, 14.5\text{Å},$ and $26.5\text{Å}$, respectively, so that the SMD pulling was between equilibrium states at the two ends of each segment.

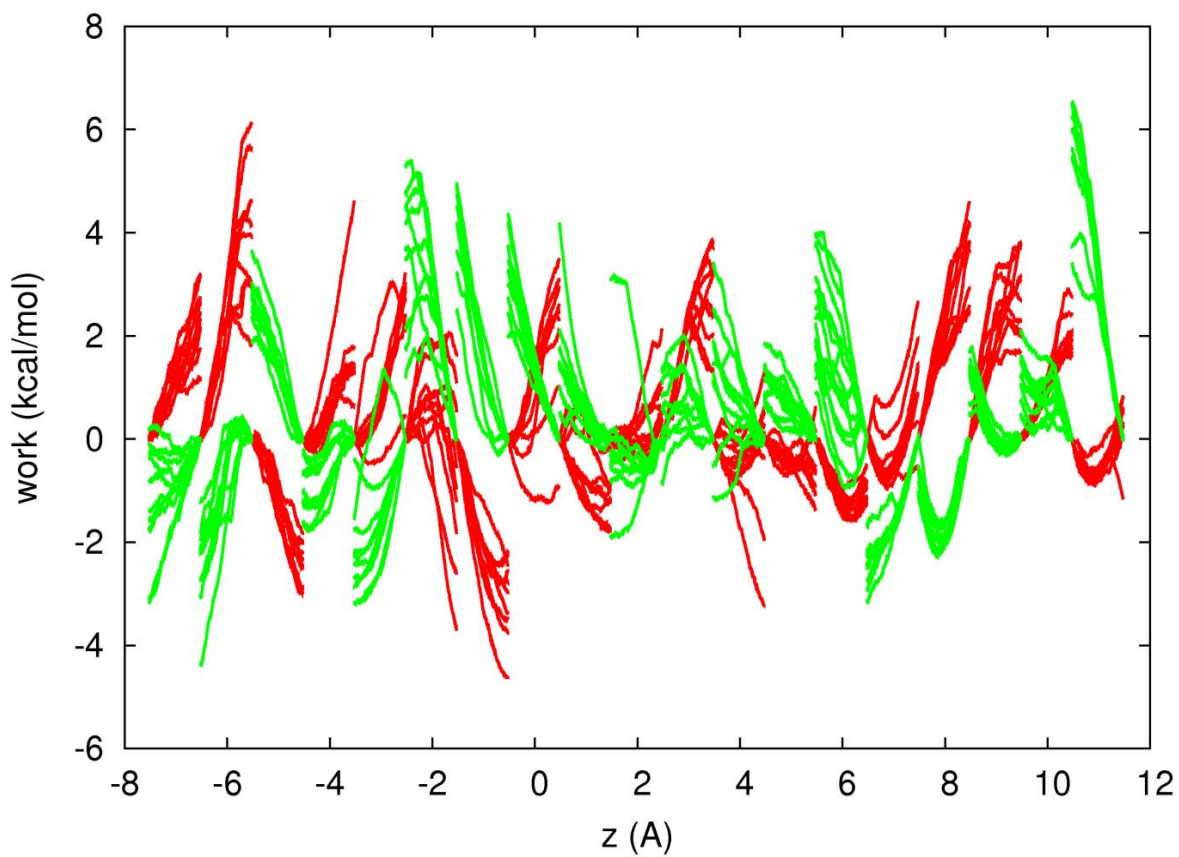

Fig. S4. Work done to the system as a function of the glycerol's center-of-mass z-coordinate along the forward (red) and reverse (green) pulling paths in SMD simulation Set #3. The conditions were identical to those of Fig. S2 except that the channel region ($-7.5\text{Å} < z < 11.5\text{Å}$) was divided into 19 segments of $1\text{Å}$ each.

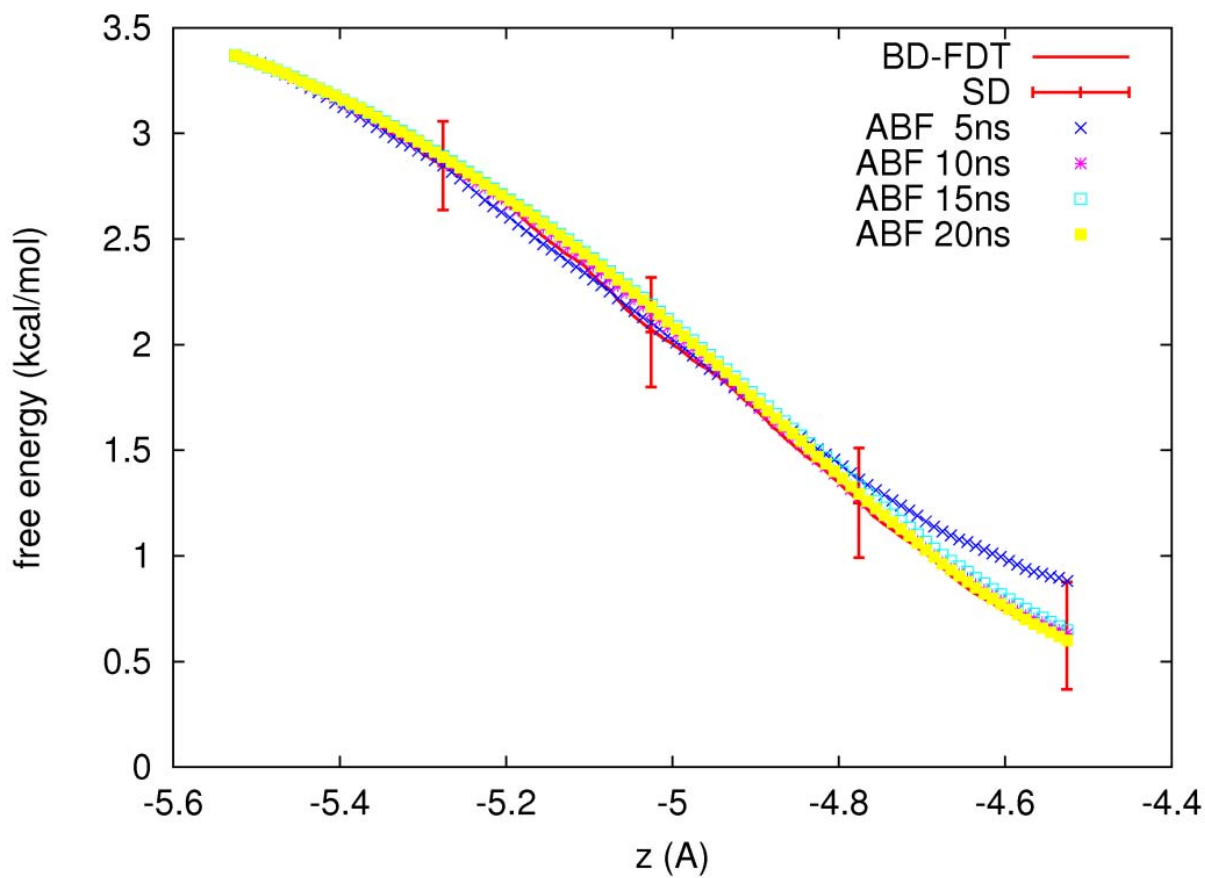

Fig. S5. Convergence of free-energy computation using the ABF method (3, 4), a mature approach of equilibrium MD. The sampling was over a window of 1Å in the z-coordinate of the center of mass of glycerol. The lengths of simulations are indicated to show the convergence behavior of the ABF approach. The BD-FDT (5) results and ABF results for 15 ns and for 20 ns overlap almost completely, suggesting the accuracy of both approaches.

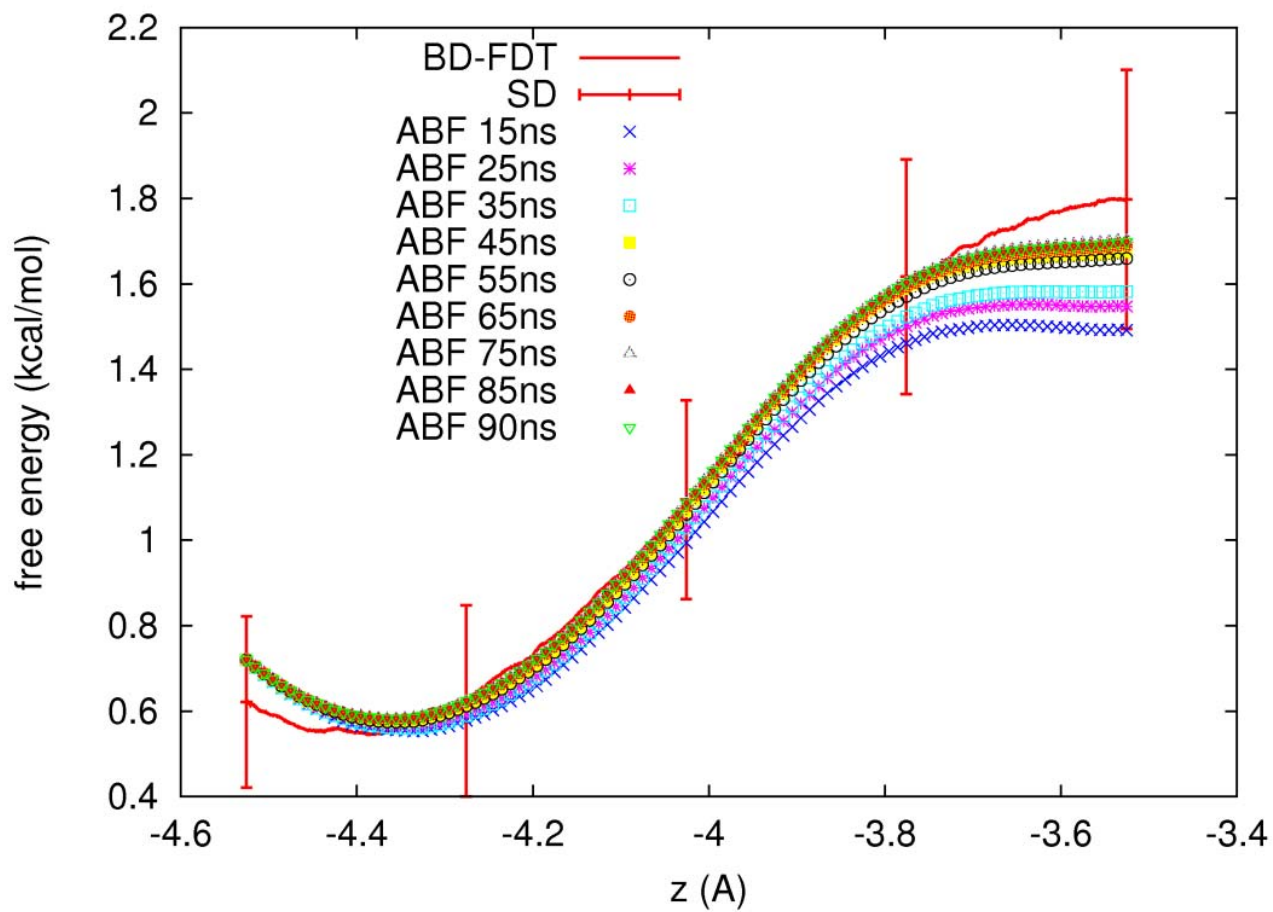

Fig. S6. Convergence of free-energy computation using the ABF method (3, 4). The lengths of simulations are indicated to show the convergence behavior of the ABF approach. The results for 75 ns, 85 ns, and 90 ns overlap almost completely and all fall within the error bars of the nonequilibrium BD-FDT approach (5), suggesting the accuracy of both approaches.

References

1. Zhu F, Tajkhorshid E, & Schulten K (2004) Collective diffusion model for water permeation through microscopic channels. *Phys Rev Lett* 93(22):224501.
2. Jensen MØ & Mouritsen OG (2006) Single-Channel Water Permeabilities of Escherichia coli Aquaporins AqpZ and GlpF. *Biophys J* 90(7):2270-2284.
3. Darve E & Pohorille A (2001) Calculating free energies using average force. *The Journal of Chemical Physics* 115(20):9169-9183.
4. Henin J & Chipot C (2004) Overcoming free energy barriers using unconstrained molecular dynamics simulations. *The Journal of Chemical Physics* 121(7):2904-2914.
5. Chen LY (2011) Exploring the free-energy landscapes of biological systems with steered molecular dynamics. *Physical Chemistry Chemical Physics* 13(13):6176-6183.